\documentstyle[12pt]{article}
\def\bendequ#1#2{\begin{equation}\label{#1}#2\end{equation}}
\def\parc#1#2{\frac{\partial #1}{\partial #2}}
\def\RE{{\rm Re}}
\def\V#1{{\bf #1}}
\def\vz{\v z}
\begin{document}
\begin{center}
{\Large \bf Bound States in the Vortex Core}\\[1.5\baselineskip]


{\large  I. Kne{\vz}evi\'c and Z. Radovi\'c}\\[1.5\baselineskip]

Department of Physics, University of Belgrade,
P.O.Box 368, 11001 Belgrade, Yugoslavia \\[.5\baselineskip]

\end{center}

\vspace{\baselineskip}
\noindent
\underline{\bf Keywords}: vortex core, 
bound states, $d$-wave superconductor

\vspace{\baselineskip}
\section*{Abstract}
{The quasiparticle excitation spectrum of isolated vortices in 
clean layered $d$-wave superconductors is calculated. A large peak in the density of states in the "pancake" vortex core is found, in an agreement with the recent experimental data for high-temperature superconductors.}

\section{Introduction}

The vortex core in classical type-II superconductors can be treated as the normal metal area with radius of the order of the coherence length $\xi(0)\sim 10\,{\rm nm}$.\cite{Dezen,Bardin} The spectrum of the bound states near the Fermi surface, formed by the constructive interference between the incident and the Andreev reflected quasiparticles,\cite{Andreev} is quasicontinuous (gapless). High-temperature superconductors (HTS)
are layered, having a cylindrical Fermi surface, 
the superconductivity is of the strong and $d$-wave coupling type, and  
the vortex core radius is much smaller, $\xi(0)\sim 1{\rm nm}$.

\vskip .5\baselineskip
For a two-dimensional (2D) vortex and $s$-pairing, Rainer {\it et al.}  have shown, using the Andreev quasiclassical theory in the analytical, and Eilenberger's in the numerical part of their study, that bound states exist in the vortex core.\cite{Rajner}
Similar conclusions have also been obtained by Maki and coworkers for $d$-pairing, within the Bogoliubov-de Gennes approach.\cite{Maki} 

\vskip .5\baselineskip
In this paper, the Eilenberger quasiclassical equations \cite{Eilenberger}, in the case of both $s$ and $d$-pairing, are solved analytically, within the  model considering the spatial variation of the order parameter in the  vortex core as for a normal metal cylinder of radius $r_c\sim \xi$. A crucial difference is found in the quasiparticle spectra below the bulk energy gap, between the classical superconductors with spherical Fermi surface, $s$-pairing, large $\xi$, and HTS with cylindrical Fermi surface, $d$-pairing, small $\xi$. Our results confirm that the quasiparticle density of states (DOS) has one large maximum, recently observed in YBCO by scanning tunneling microscopy.\cite{Magio} 

\section{Model of the Vortex Core}

An efficient method for calculating local spectral properties is the quasiclassical theory of superconductivity, which gives the Eilenberger equations 
\bendequ{E4}
{\left[ 2\hbar\omega\sb n +\hbar \V v\cdot\left(\V \nabla -\imath \frac
{2e}{\hbar c}\V A\right)\right] f=
2\Delta g,\quad \left[ 2\hbar\omega\sb n -\hbar \V v\cdot\left(\V \nabla +\imath\frac
{2e}{\hbar c}\V A\right)\right] f\sp{\dagger}=
2\Delta\sp * g, }
\bendequ{E6}
{\hbar\V v\cdot\V \nabla g=\Delta\sp *f-f\sp{\dagger}\Delta .}
Here $g=g_{\downarrow\downarrow}(\V r,\V v,\omega\sb n)$ and $f=f_{\downarrow\uparrow}(\V r,\V v,\omega\sb n)$ represent the normal and the anomalous Green function  respectively, $\Delta =\Delta (\V r,\V v)$ is the gap function, $\omega\sb n=\pi k\sb B T(2n+1)$, $n=0,\pm 1,\pm 2...$, 
are the Matsubara's frequencies, and $\V v$ is the  Fermi velocity vector. The function  $f\sp{\dagger}$ is defined by $f_{\uparrow\downarrow}\sp \dagger(\V r,\V v,\omega\sb n)=f_{\downarrow\uparrow}\sp *(\V r,-\V v,\omega\sb n)$. $\Delta$ and $f$ are connected by the self-consistency equation. 

\vskip .5\baselineskip
For a homogeneous and isotropic superconductor, solutions of 
the Eilenberger equations are
\bendequ{E3a}
{\langle f\rangle=\frac{\Delta}{\varepsilon_n},
\quad \langle f\sp{\dagger}\rangle=
\frac{\Delta\sp*}{\varepsilon_n},
\quad \langle g\rangle =\frac{\hbar \omega\sb n}{\varepsilon_n}\qquad 
\left(\varepsilon_n\sp 2=\vert \Delta\vert\sp 2+(\hbar\omega\sb n)\sp 2\right).}

In the cylindrical coordinates $(r,\varphi ,z)$, with the origin situated on the vortex axis, the vortex magnetic field is $\V h=h\V e_z$. 
At distances  $r\sim \xi\ll\lambda$ from the vortex axis, $h$ is approximately constant, and the gauge can be chosen in the form $\V A=\left(\V h\times \V r\right)/2$. Taking the same vortex gap function as for a normal metal cylinder embedded in a superconductor
\bendequ{E8}
{\Delta=\Delta(r,\theta)e\sp{-\imath\varphi}\ ,\ \Delta(r,\theta)=
\cases{0,& $r\leq r_c$\cr \Delta(\theta),& $r > r_c$}\qquad .}
Here, $r_c$ is the vortex core radius, and $\theta$ is the polar angle in $\V k$-space. 
For $d$-pairing $\Delta (\theta)=\Delta_0\cos 2\theta$, and for $s$-pairing $\Delta (\theta) =\Delta_0$. The gauge in Eq. (\ref{E8}) is due to the flux quantization. 

\vskip .5\baselineskip
For a pancake vortex in ($r,\varphi $) plane, 
denoting the coordinate along $\V v$ by $s$, and along  
$\V h\times \V v$ by $p$ (Fig.1.), 
in the gauge with real gap, Eqs. (\ref{E4}) and (\ref{E6}) can be rewritten 
in the form
\bendequ{E13}
{\left[ 2\hbar\omega\sb n +\hbar {\rm v}\left(\parc{}{s} +
\imath\frac{p}{l\sb H\sp 2}+\imath\frac{p}{r\sp 2}\right)\right] f=2\Delta (r,\theta)g,}
\bendequ{E13a}
{\left[ 2\hbar\omega\sb n -\hbar  {\rm v}\left
(\parc{}{s} -
\imath\frac{p}{l\sb H\sp 2}-\imath\frac{p}{r\sp 2} \right)\right] f\sp{\dagger}=
2\Delta(r,\theta) g,}
\bendequ{E15}
{\hbar {\rm v}\parc{}{s}g=\Delta(r,\theta)(f-f\sp{\dagger}),}
where $r^2 =p^2 +s^2$ and $l\sb H\sp 2=\hbar c/eh$. 

\vskip .5\baselineskip
For a normal metal cylinder and zero magnetic field, 
Eqs. (\ref{E13})-(\ref{E15}) with 
$p=0$, the solution is of the form 
\bendequ{E16}
{f=\sum\sb i  f\sb i(p)e\sp{\kappa\sb i s},\quad  
g=\sum\sb i  g\sb i(p)e\sp{\kappa\sb i s}.} 
For $r\leq r_c$, with $\kappa\sb 0=2\omega_n/{\rm v} $,
\bendequ{E18}
{f=Fe\sp{-\kappa\sb 0 s},\quad  g=G.}
For $r> r_c$ and  $\kappa=2\varepsilon_n/\hbar {\rm v}$, 
\bendequ{E19}
{f=\langle f\rangle +\Phi\sb 1 e\sp{-\kappa s},
\quad g=\langle g\rangle +\Gamma\sb 1 e\sp{-\kappa s},\quad\mbox{for $s>0$},}
\bendequ{E20a}
{f=\langle f\rangle +\Phi\sb 2 e\sp{\kappa s}
,\quad g=\langle g\rangle +\Gamma\sb 2 e\sp{\kappa s},\quad\mbox{for $s\leq 0$.}}
Eqs. (\ref{E13})-(\ref{E15}) imply
\bendequ{E21}
{\frac{\Phi\sb 1}{\Gamma\sb 1}=\frac{\Delta (\theta)}{\hbar \omega_n -\varepsilon_n},\quad  
\frac{\Phi\sb 2}{\Gamma\sb 2}=\frac{\Delta (\theta)}{\hbar \omega_n +\varepsilon_n}.}
Using the continuity condition for $f$ and $g$ at 
$\pm s\sb 0=\pm \sqrt{r_c^2-p^2}$, for $r<r_c$ the normal Green function is 
\bendequ{E25}
{G=\frac{\hbar\omega_n\cosh (\kappa_0 s_0)+\varepsilon_n \sinh (\kappa_0 s_0)}
{\hbar\omega_n\sinh (\kappa_0 s_0)+\varepsilon_n \cosh (\kappa_0 s_0)}.}
For a vortex, approximating $p/r^2$ by $p/r_c^2$, 
the solution of Eq. (\ref{E13})-(\ref{E15}) 
can be obtained from Eq. (\ref{E25}), by changing 
$\omega_n\to{\omega_n}'$, 
\bendequ{Gp}
{G\approx \frac{\hbar\omega_n '\cosh (\kappa_0 's_0)+\varepsilon_n '\sinh (\kappa_0 's_0)}
{\hbar\omega_n '\sinh (\kappa_0 's_0)+\varepsilon_n '\cosh (\kappa_0 's_0)},}
where 
\bendequ{E25a}
{{\omega_n}'=\omega_n +\imath\frac{p{\rm v}}{2}\left(\frac{1}{r_c^2}+\frac{1}{l_H^2}\right),}
and $\kappa_0 '=2\omega_n '/{\rm v}$. 
In this case, the magnetic flux quantization leads to
\cite{Dezen,Maki}
\bendequ{E26}
{p_i  =\left( i+\frac{1}{2}\right)\frac{\hbar}{m{\rm v}} ,\quad i=0,\pm 1,\pm 2,...}
Since for an isolated vortex $l_H\gg r_c$, the direct influence of the field can be neglected,  and the only relevant contribution is due to the screening supercurrent flow, $\imath p{\rm v}/2r_c^2$ term in Eq. (\ref{E25a}). 

\section{Bound States}

Performing an analytical continuation of $G$ by $\hbar\omega_n\to -\imath E +\eta$, $E$ being the quasiparticle energy with respect to the Fermi level, the retarded propagator $g^R(E, p, \theta )$ is obtained.  In terms of  reduced variables $E/\Delta_0\to E$,
$\sqrt{\Delta^2(\theta)-E^2}/\Delta_0 \to \varepsilon$, $\sqrt{E^2 -\Delta^2(\theta) }/\Delta_0 \to e $,  $p/r_c\to p$, $2s_0/\pi \xi_0 \to s_0$, $\xi_0=\hbar{\rm v}/\pi\Delta_0$ being the BCS coherence length, 
angle resolved partial DOS (PDOS) is obtained from $N(E, p, \theta )=\RE g^R(E, p, \theta )$. 

\vskip .5\baselineskip
For the normal metal cylinder
\begin{eqnarray}\label{E30}
N(E , p, \theta )/N(0)&=&\Theta \left(e ^2\right)
\frac{|E|e }{e ^2\cos^2 (Es_0)
+E^2\sin^2 (Es_0)}+\nonumber\\
&+&\Theta \left(\varepsilon^2\right)\frac{\pi |\Delta(\theta)|}{\Delta_0}\delta \left(E\sin (Es_0)-
\varepsilon\cos (E s_0)\right),\end{eqnarray}
where $\delta$ is the Dirac function, $\Theta$ is the step-function, and $N(0)=m/2\pi\hbar^2$ 
is the normal metal density of states at the Fermi surface for one spin orientation. For $s$-wave pairing, 
PDOS does not depend on $\theta$, while for $d$-wave pairing, 
averaging over the cylindrical Fermi surface leads to
\begin{eqnarray}\label{E39}
N(E,p )/N(0)&=&\frac{1}{2\pi}\int_0^{2\pi} N(E, p, \theta )/N(0)\, d\theta =\nonumber\\
&=&\frac{2}{\pi}\int_{\sqrt{\max \{ 0, E^2 -1\}}}^{|E|}\, 
\frac{|E|e ^2\, de }{\sqrt{E^2-e ^2}
\sqrt{1-E^2+e ^2}
\left(e ^2\cos^2(E s_0)+
E^2\sin^2(E s_0)\right)}+\nonumber\\
&+&\frac{\left(E \tan (E s_0)+|E \tan (E s_0)|\right)}
{\sqrt{\cos^2 (E s_0)-E^2}}
\Theta \left(\cos^2 (E s_0)-E^2\right). \end{eqnarray}
Finally, after spatial averaging over the cylinder area $\pi r_c^2$, DOS is 
\bendequ{E40}
{N(E)=\frac{4}{\pi}\int_0^1 N(E,p)\sqrt{1-p^2} \, dp.}
For the vortex, in Eqs. (\ref{E30}) and (\ref{E39}), $E\to E+E_0\,{\rm sign}\, E$, 
$E_0={\hbar |p_i|{\rm v}}/{2\Delta_0r_c^{2}}$, Eq. (\ref{E25a}), 
with $p_i$ from Eq. (\ref{E26}), and $\sum_{p_i}$ instead of integration in Eq. (\ref{E40}). Here, signs of $p$ and $E$ are connected, because the magnetic field causes a difference in propagation of particles and holes. 

For small radius vortices in HTS, $\xi_0 m{\rm v}/\hbar\sim 1$ ($\sim 10$ in classical superconductors), only one trajectory through the vortex core is allowed, with $p=p_0$, Eq. (\ref{E26}). 
Taking for YBCO $r_c =\xi_0$, $p_0=1/3$, $\Delta_0/E_F=0.424$, only one peak 
in DOS in the vortex core around $E/\Delta_0\approx 0.3$ is obtained (Fig. 2). For comparison, DOS of normal metal cylinder embedded in the same superconductor and with the same radius $r_c=\xi_0$, but in the zero magnetic field, is shown. In this case, a large energy gap is found in DOS, due to formation of lowest bound state at high energy, of the order of $\Delta_0$. This is not the case in classical superconductors, where $\Delta_0/E_F\ll 1$. 

\vskip \baselineskip
\begin{minipage}[h]{6cm}{
\vskip 4cm
{\bf Fig.1.} Trajectory passing at distance $p$ from the vortex center.\label{vort}}
\end{minipage}
\hspace{1cm}
\begin{minipage}[h]{10cm}{
\vskip 7cm
{\bf Fig.2.} Quasiparticle DOS in the vortex core (solid curve), and in the normal metal cylinder (dashed curve), for 
clean layered $d$-wave superconductor. $r_c/\xi_0 =1$, $\Delta_0/E_F =0.424$.\label{Grafik}}
\end{minipage}

\vskip \baselineskip
In conclusion, cylindrical Fermi surface, $d$-wave pairing and small $\xi_0$, 
large $\Delta_0/E_F\sim 0.1$, make DOS of a normal metal cylinder embedded in HTS and a 
pancake vortex different from DOS of a normal cylinder and a vortex in classical superconductors. 
Since the Andreev bound states can transport charge currents, unlike the bound states in a potential well, 
supercurrents can flow through the vortex without losses, strongly influencing its dynamics. 
This could be very important for transport properties of HTS, especially for understanding the unusual  magnetic-field dependence of the electrothermal conductivity, which was observed experimentally  and awaits explanation.\cite{Bozovic}


\begin{thebibliography}{99}
\bibitem{Dezen}
C. Caroli, P. G. de Gennes and J. Matricon, Phys. Lett. {\bf 9}, 307 (1964).
\vspace{-.5\baselineskip}
\bibitem{Bardin}
J. Bardeen and M. J. Stephen, Phys. Rev. {\bf 140}, 1197 (1965).; {\it ibid} {\bf 187}, 
556 (1969).
\vspace{-.5\baselineskip}
\bibitem{Andreev}
A. F. Andreev, Sov. Phys. JETP {\bf 19}, 1228 (1964). 
\vspace{-.5\baselineskip}
\bibitem{Rajner}
D. Rainer, J. A. Sauls and D. Waxman, Phys. Rev. B {\bf 54}, 10094 (1996).  
\vspace{-.5\baselineskip}
\bibitem{Maki}
N. Schopohl and K. Maki, Phys. Rev. B {\bf 52}, 490 (1995).; Y. Morita, M. Kohmoto and K. Maki, Phys. Rev. Lett. {\bf 78}, 4841 (1997).
\vspace{-.5\baselineskip}
\bibitem{Eilenberger}
G. Eilenberger, Z. Phys. {\bf 190}, 142 (1966); {\em ibid} {\bf 214}, 195 (1968).
\vspace{-.5\baselineskip}
\bibitem{Magio}
I.Maggio-Aprile {\it et al.}, Phys. Rev. Lett. {\bf 75}, 2754 (1995).
\vspace{-.5\baselineskip}
\bibitem{Bozovic}
J. A. Clayhold, Y. Y. Xue, C. W. Chu, J. N. Eckstein and I. Bozovic, Phys. Rev. Lett. {\bf 53}, 8681 (1996).

\end{thebibliography}
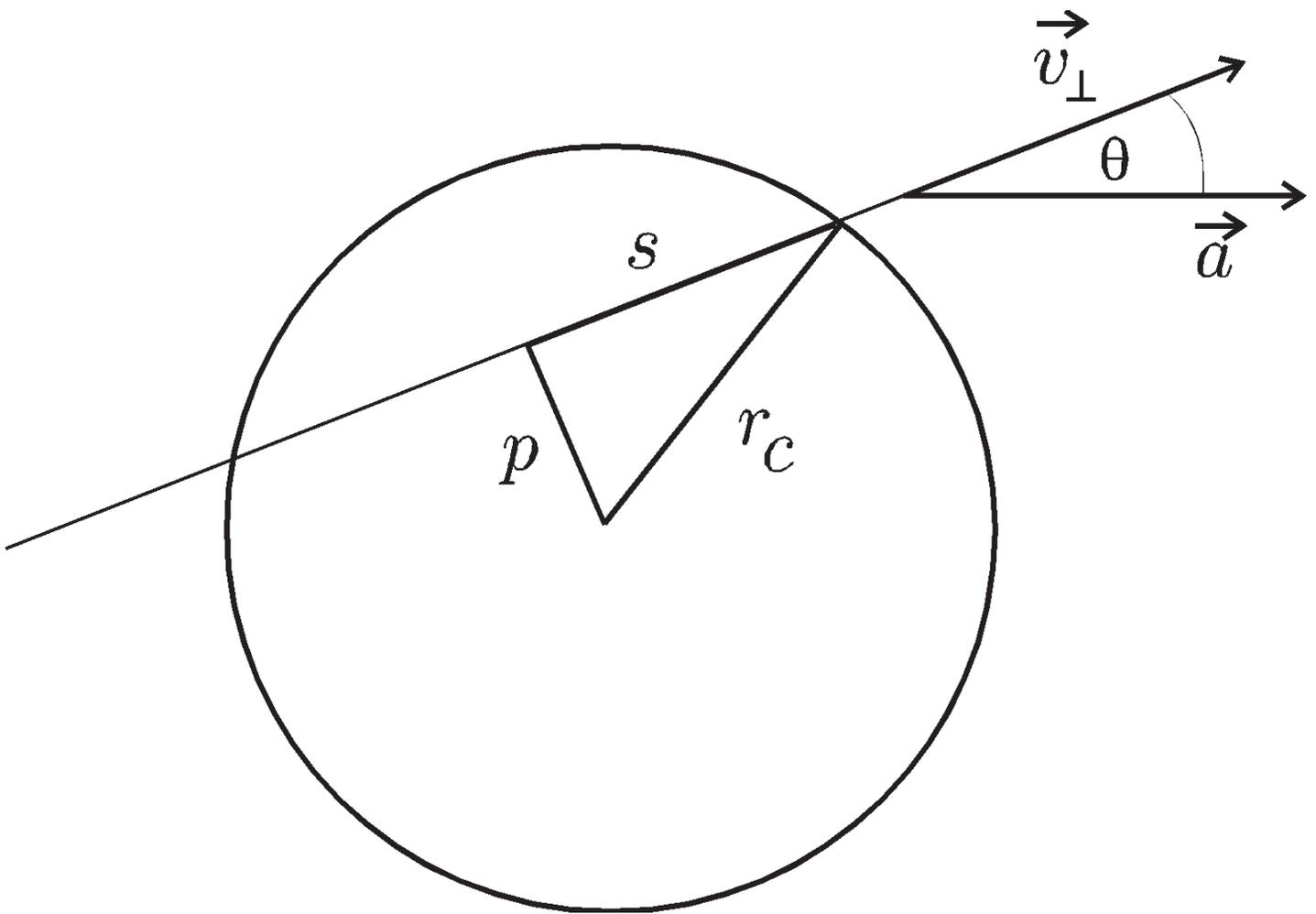
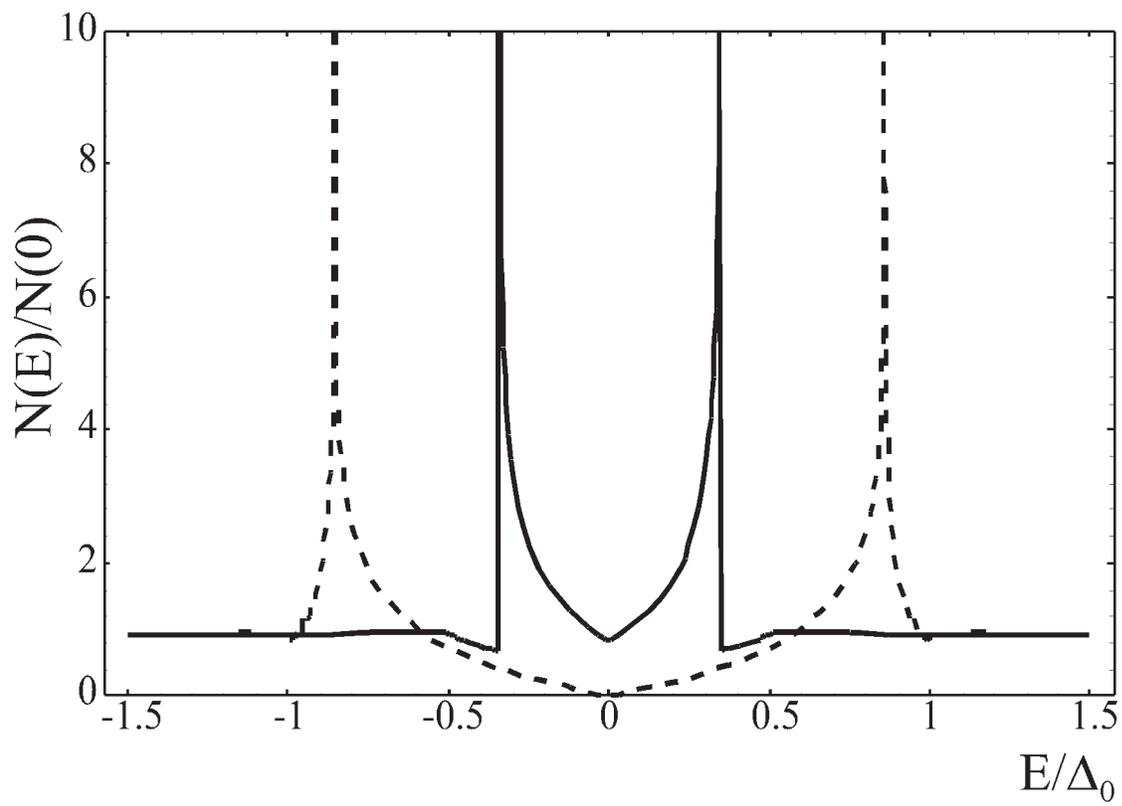
\end{document}